# Challenges in Automatic Speech Recognition for Adults with Cognitive Impairment

Michelle Cohn, Alyssa Lanzi, Yui Ishihara, Chen-Nee Chuah, Georgia Zellou, Alyssa Weakley

**Abstract:**
Millions of people live with cognitive impairment from Alzheimer's disease and related dementias (ADRD). Voice-enabled smart home systems offer promise for supporting daily living but rely on automatic speech recognition (ASR) to transcribe their speech to text. Prior work has shown reduced ASR performance for adults with cognitive impairment; however, the acoustic factors underlying these disparities remain poorly understood. This paper evaluates ASR performance for 83 older adults across cognitive groups (cognitively normal, mild cognitive impairment, dementia) reading commands to a voice assistant (Amazon Alexa). Results show that ASR errors are significantly higher for individuals with dementia, revealing a critical usability gap. To better understand these disparities, we conducted an acoustic analysis of speech features and found that a speaker's intensity, voice quality, and pause ratio predicted ASR accuracy. Based on these findings, we outline HCI design implications for AgeTech and voice interfaces, including speaker-personalized ASR, human-in-the-loop correction of ASR transcripts, and interaction-level personalization to support ability-based adaptation.

which suggests that fine-tuning ASR models with these features could lead to improved accessibility.

**Introduction.**
Millions of people currently have Alzheimer's disease and related dementias (ADRD) and this number is increasing, with 13.8 million cases anticipated by 2060 in the United States [101]. There is a rapid development of systems to support aging-in-place for older adults with cognitive impairment (AgeTech), such as smart home systems, home helper robots, and voice assistants [26, 47, 60, 66, 81, 84, 92, 100]. AgeTech systems can screen for mild cognitive impairment (MCI), a precursor to ADRD for many individuals, such as based on language [46, 90] and movement [51]. AgeTech can also support activities of daily living (ADL) [92] and social interaction [72]. Many of these systems use voice assistants, which can provide an accessible interface that does not require expertise with mobile and computer interfaces [37, 67].

In systems using spoken interaction (e.g., voice assistants, home health robots), the technology transcribes the user's speech to text via automatic speech recognition (ASR). While ASR systems continue to improve, they still often show worse transcription for speech produced by older adults [61, 87, 94], speakers with non-"standard" accents and dialects [14, 21, 38], and individuals with neurological symptoms that shape language (e.g., dysarthria [70], Down Syndrome [15]). Compared to cognitively normal controls, individuals with cognitive impairment produce acoustic differences in their speech: a slower rate, reduced intensity, lower and less variable pitch, hoarser voice quality, and more frequent pauses [3, 5, 28, 54, 58], which could impact ASR.

While there is a large body of work classifying cognitive status from speech (for a review, see Saeedi et al., 2024), few report the error rates for ASR and those that do exclusively do so as a proposed feature in detecting MCI/dementia [44, 46, 48, 93, 98, 99]. Accordingly, we know little about *when* and *why* ASR errors occur for adults with cognitive impairment.

Understanding these errors has important implications for Human Computer Interaction (HCI) frameworks. First, ASR errors meaningfully shape users' interactions with voice-based AgeTech systems, impacting trust [9], satisfaction [59], and adoption [45]. Early misrecognition can have lasting impacts: if a person experiences errors from a system early on, they are less likely to use it -- even if it later improves [16, 50]. Second, disparities in ASR based on cognitive status can speak to **Ability-Based Design (ABD)** [95]*,* which argues that technologies should adapt to users' abilities, rather than requiring them to compensate for system limitations. In the context of cognitive impairment, this includes accommodating known speech production changes associated with decline. An ability-based perspective underscores the need to understand when and why ASR disparities arise to design voice-based AgeTech that is accessible and equitable.

This paper provides the following contributions:
- **Empirical evidence** that a state-of-the-art ASR model (Whisper) shows lower transcription accuracy for older adults with dementia.
- **Acoustic analysis** demonstrating that speech characteristics associated with cognitive impairment (e.g., reduced intensity, increased shimmer) result in higher ASR error rates.
- **Design implications** for inclusive AgeTech for cognitive impairment, including personalizing ASR models, human-in-the-loop support from users and caregivers, and interaction-level adaptation informed by dementia-related speech differences (e.g., extended turn taking).

**Background and Related Work**
**AgeTech and Conversational AI**
There is a growing body of research developing systems to support older adults to maintain their independence, particularly in the face of cognitive decline, including smart-home monitoring platforms [26, 81, 100], web-based caregiving and assistive technologies [20, 65, 92], and companion robots that provide physical or social support [13, 47, 84]. For example, studies have investigated smart homes utilizing sensor-based monitoring systems that detect human activities, identify anomalies in daily routines, and even provide reminders or alerts to external caregivers [51, 74].

Other work has examined helper robots that can provide personalized physical support for older adults such as assisting with household tasks, mobility assistance, facilitating physical therapy, and reminders for medication routines (for reviews, see Asgharian et al., 2022; Sather et al., 2021). Conversational AI is increasingly used in these systems, enabling voice-based interaction between users and robots [47]. For example, HCI research has shown that robots can provide social assistance including companionship and cognitive stimulation through speech-based interactions [10].

Many studies have also investigated the role of voice assistants (e.g., Apple's Siri, Amazon Alexa, Google Assistant) in serving as AgeTech [20, 39, 56, 68, 72, 85]. For example, [56] found that individuals with dementia felt that voice assistant systems helped them cope with loneliness and improved social connectedness. Other groups have used wearable voice assistants to help older adults manage reminders; for example, the I-Wear system uses voice commands to an Apple Watch to record electronic post-it notes for adults with MCI [65]. Older adults, in particular, tend to prefer voice-based interactions with technology rather than typed interactions [76], making this an accessible mode of interaction.

Despite these opportunities to support older adults with cognitive impairment, there are barriers for them to use voice technology. For example, voice interfaces often have a set window to accept responses, which can result in the assistant interrupting the user if they are speaking at a slow rate, such as with dementia [1]. Additionally, individuals with cognitive impairment show speech differences from age-matched controls [3, 28, 54, 58], which could have downstream effects on how well voice technology understands them

**Barriers to inclusive ASR**
Voice interfaces typically use ASR to transcribe a user's speech into text; this text can then serve as the input to a natural language understanding (NLU) model or a large language model (LLM) (but see [91] for speech-to-speech interfaces). Prior work has shown that ASR systems show disparities in the varieties of language they can accurately transcribe. In particular, ASR produces less accurate transcriptions based on speakers' social characteristics including age [61, 94], gender [25, 38], race/ethnicity [38], region [25], and first language [14, 88]. For example, ASR accuracy is often lower for older adults [94] and children [75], relative to college- and middle-age adults. Furthermore, accuracy is often higher for women relative to men [25, 38, 94]). Additionally, ASR accuracy has shown to be lower for sociolects (i.e., language varieties associated with race), such as lower accuracy for African American English speakers than white 'mainstream' English speakers [38]. In addition to biases by social characteristics, there are worse ASR outcomes for speakers who have clinical sources for their speech differences, such as dysarthria [70], Down Syndrome [15], Parkinson's Disease [62], Primary Progressive Apraxia of Speech [79], and cognitive impairment [46, 77]. Reduced ASR accuracy can have downstream effects on how technology can understand people [21], which can lead to frustration and feelings of being "othered" [59], as well as abandonment of the technology [57].

In terms of cognitive impairment, several papers have reported reduced accuracy [46, 77]. For example, [77] found that compared cognitively normal older adults, individuals with memory complaints, mild cognitive impairment, and dementia combined had a higher word error rate (WER) for ASR on reading and spontaneous speech tasks (CN = 0.15-0.34; combined cognitively impaired group = 0.23-0.34). Similarly, [46] found that cognitively normal older adults had lower WER (0.22) than those with MCI (0.27) on read voice assistant commands.

Of note, the majority of prior work investigating ASR performance for individuals with cognitive impairment has examined it as a potential input to a classifier distinguishing normal and impaired cognition [44, 46, 48, 93, 98, 99]. For example, [44] found that dementia detection models performed better when they incorporated transcripts containing ASR errors, compared to using error-free ground-truth transcripts. This further suggests disparities in ASR errors for cognitively impaired older adults. However, little prior work has attempted to understand the source of these differences and how to adapt ASR models to improve interaction.

**Novelty, Contribution, & Research Questions (RQ)**
The current study consists of several novel contributions. First, we conducted a quantitative analysis of ASR transcription of speech produced by older adults with MCI and dementia. While prior work has statistically evaluated ASR disparities with mixed-effects models, which take into account speaker- and item-level variation and other demographic factors (e.g., age, gender, language variety, etc.), for other clinical disorders [82] and language varieties [25, 88], research on cognitively normal adults and those with MCI/dementia has been limited to descriptive statistics [46, 77] and ASR performance as an input for a machine learning classifier to detect cognitive impairment [44, 46, 48, 93, 98, 99]. Consequently, ASR performance gaps between adults with and without cognitive impairment remain poorly understood, limiting our ability to both identify differences and quantify future improvements in ASR accuracy.

- RQ1: Is ASR performance lower for individuals with cognitive impairment than cognitively normal older adults? We used two Whisper ASR models to assess transcriptions to determine whether biases against speech produced by individuals with cognitive impairment are consistent. We predict that accuracy will be highest for adults with no cognitive impairment, lower for folks with MCI, and lowest for individuals with dementia, stemming from the past papers reporting descriptive statistics [46, 77].

- RQ2: Is ASR performance predicted by speech differences produced by the groups? We provide an acoustic analysis, examining timing, voice quality, and prosodic features previously identified as potential 'biomarkers' of cognitive impairment. The current study is the first quantitative analysis of ASR performance linked to acoustic features across individuals with varying cognitive status. We hypothesize that the features distinguishing the speech of cognitively normal adults from those with impairment will also predict ASR accuracy.

**Methods**
**Dataset: VAS Corpus**
We analyzed ASR transcription of Voice Assistant System (VAS) corpus [42, 46] available through DementiaBank [43]. The dataset consists of 101 older adults who read voice commands to an Amazon Echo either in-person (n=11) or virtually (n=90; during the Covid-19 pandemic).

Speakers read a fixed set of 30 Alexa voice commands (see **Appendix A**). These included internet-of-things commands (e.g., "Alexa, turn on the bedroom light on."), reminders (e.g., "Alexa, remember my daughter's birthday is June first."), and questions (e.g., "Alexa, what time is it?"). Speakers read the commands in the same order and engaged a real system through Zoom (see [46] for details). Audio recordings of the voice commands in VAS come from the Alexa app. All audio recordings were transcribed by human annotators using the CHAT format [52].

**Speakers**
The full sample includes 30 individuals with dementia, 35 with MCI, and 36 cognitively normal older adults. Group membership was determined by clinical diagnosis (for individuals with dementia) or based on their score on the Montreal Cognitive Assessment (MoCA) [63]. Individuals scoring between 26-30 were classified as 'cognitively normal' and those with scores from 20-25 were classified as having mild cognitive impairment (MCI). Individuals who had a clinical diagnoses of dementia who scored greater than 19 were retained in the dementia category. No individuals were clinically classified as either MCI or CN. All speakers were English-speaking and living in the United States. Speakers completed informed consent through Dartmouth and University of North Carolina (UNC) Institutional Review Boards (IRBs) and secondary analysis was approved by the University of California, Davis IRB.

**Speakers Included in the Current Analysis**
Our analyses focused on a subset of the full VAS corpus. We excluded speakers who had MoCA scores indicative of severe cognitive impairment (< 10; n=4). These speakers also struggled with the task (e.g., not able to read the full target sentences aloud, leading to incomplete recordings; e.g., P84 only had 3 complete commands, P074 had 2 complete commands). We also limited our analysis to speakers who completed the study virtually as only 11 speakers completed the study in-person, and they were not balanced by cognitive status. Two speakers in the MCI group with MoCA scores under 20 were also removed. Retained speakers (n=83) are summarized in **Table 1** below.

**Table 1**. Speaker demographics in the current study

|  | Cognitively normal (n=30) | Mild cognitive impairment (MCI) (n=28) | Dementia (n=25) |
| --- | --- | --- | --- |
| Age | mean = 70.93 years (0.67) | mean = 72.21 years (0.82) | mean = 77.84 years (1.74)*** |
| Gender | 13 female, 17 male | 15 female, 13, male | 14 female, 11 male |
| Race | 1 Asian, 2 Black/African American, 27 white | 1 Asian, 0 Black/African American, 27 white | 0 Asian, 5 Black/African American, 20 white |
| Highest Education | 1 high school; 2 some college; 10 college degree; 17 post-college degree | 0 high school; 8 some college; 8 college degree; 12 post-college degree | 1 high school; 7 some college; 6 college degree; 11 post-college degree |
| Voice assistant experience | 17 yes, 13 no | 12 yes, 16 no | 9 yes, 16 no |

Significant differences from cognitively normal are indicated  (* = $p<0.05$, ** = $p<0.01$, *** = $p<0.001$).

## Analysis & Results
### ASR Accuracy

We segmented each utterance based on the CHAT timestamps. We analyzed a subset of commands (n=29), excluding the phone call command ("Alexa, call (603)660-2203.") which had many errors for all speakers.

Next, we matched the 'ground truth' human annotated transcript with the intended command (see **Appendix A**). We excluded productions that were off-task (e.g., talking to the experimenter). We then ran each utterance once through a local instance of Whisper (REF) 'small' model (244 M parameters) and 'medium' model (769 M parameters) with Python (Version 3.8) (see **Appendix B** for more details).

While larger Whisper models are available, both small and medium models can be run on a laptop locally, which is privacy preserving for HIPAA. We compared two models to assess whether differences on the basis of cognitive status are consistent across models with more parameters. We then cleaned up both the ground truth and ASR transcription, removing punctuation, converting to lowercase, and converting numbers to English with the *english* R package [23]. Next, we ran the cleaned ASR and ground truth transcripts with the *JiWER* Python package [86] to extract the word error rate (WER). WER is based on the number of errors (substitutions, deletions, insertions) relative to the number of words in the ground truth (i.e., human annotation).

We modeled WER (centered) with a mixed effects logistic regression with the *lme4* R package [8]. Fixed effects included Cognitive Status (cognitively normal (CN), MCI, dementia; treatment coded, ref = CN), Command Category (Internet of Things (IoT), reminder, question; sum coded), ASR Model (small, medium; sum coded) and all possible two-way interactions (note that including 3-way interactions led to collinearity). Fixed effects also included Proportion of Experiment (centered) and speaker demographics: Age (centered), Gender (female, male; sum coded), Race (white, Black/African American, Asian; sum coded), Education (less than high school < high school degree < some college < college degree < professional degree; ordered factor), and Prior Voice Technology Experience (yes, no; sum coded). Random effects included by-Speaker and by-Trial random intercepts, and by-Speaker random slopes for Command Category and Proportion of Experiment. In the case of singularity or convergence errors, we simplified the random effects structure [6]. We assessed collinearity with the *performance* R package [49].

Predictors in the WER model had low collinearity (all VIF < 5) and thus all could be retained. The model (output provided in **Appendix C**) showed effects of Cognitive Status: individuals with dementia had higher WER than cognitively normal adults (as seen in **Figure 1**) (*Coef* = 0.14, *t* = 5.17, *p* < 0.001). There was no difference between cognitively normal adults and adults with MCI (*p* = 0.87). There was also a difference by ASR Model: WER was lower for the medium model (*Coef* = -0.03, *t* = - 6.04, *p* < 0.001). There were also interactions between Cognitive Status and Command Category: as seen in **Figure 1,** adults with dementia had higher WER for IoT commands (*Coef* = 0.11, *t* = 7.70, *p* < 0.001), but lower for Reminders (*Coef* = -0.09, *t* = -6.76, *p* < 0.001). There was also an interaction between Cognitive Status and ASR Model: the higher WER for dementia was reduced with the 'medium' model (*Coef* = -0.08, *t* = -8.84, *p* < 0.001). Finally, there were interactions between Command Category and ASR model: the medium model had a reduced WER for IoT commands (*Coef* = -0.01, *t* = -2.78, *p* < 0.01) and Questions (*Coef* = -0.01, *t* = -2.01, *p* = < 0.05) but less of a decrease in WER for Reminders (*Coef* = 0.02, *t* = 4.61, *p* < 0.001). No other effects or interactions were observed.

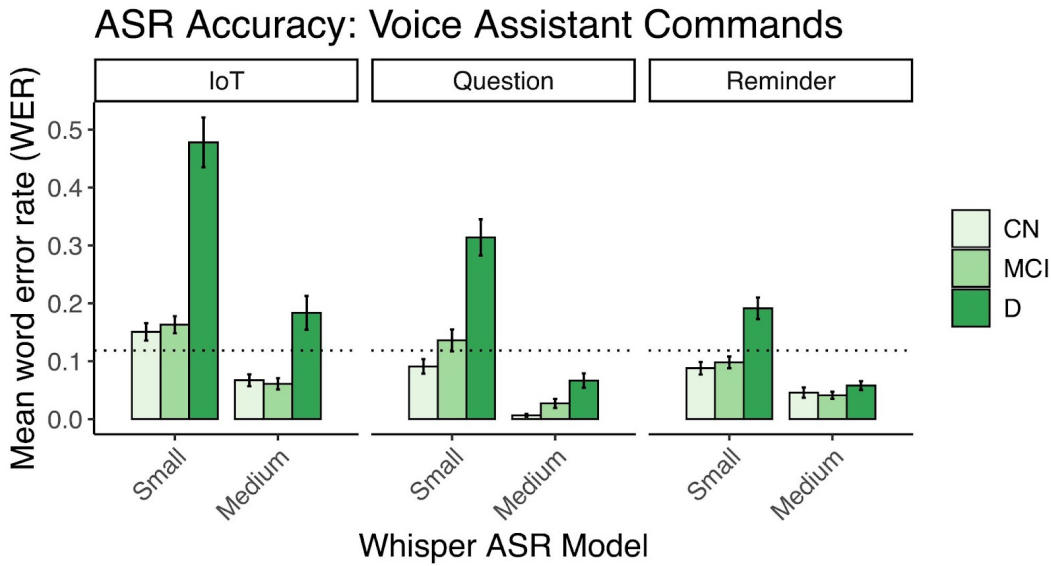

**Figure 1**. Automatic speech recognition (ASR) accuracy for voice assistant commands in the VAS corpus [46] across adults who were cognitively normal (CN), had mild cognitive impairment (MCI), or had dementia (D). Accuracy, measured in word error rate (WER), was assessed across three command categories: Internet of Things (IoT), Questions, and Reminders. The x-axis shows the two Whisper ASR models compared: 'small' and 'medium' English models. The dotted line shows the grand mean WER, while error bars indicate standard error of the mean.

**Acoustic Analysis & Results**
We extracted utterance-level measurements using Praat [11] across three categories: timing, voice quality, and prosody. **Timing** measures included speech rate (number of syllables/second), articulation rate (number of syllables/second excluding silence), and silent pause ratio (duration silence/total duration) [35]. **Voice Quality** measures included jitter (cycle-to-cycle variability in pitch) and shimmer (cycle-to-cycle variability in intensity). **Prosodic measures** included intensity, mean pitch and pitch variability. For pitch measurements, fundamental frequency (f0, perceived as pitch) was sampled at 10 equidistant points over the utterance [19] and converted to semitones to be on a perceptually linear scale before taking the mean and standard deviation.

Our first analysis was to test whether the acoustic features we measured are associated with a speaker's cognitive status. We modeled each feature – speech rate, articulation rate, pause ratio, intensity, jitter, shimmer, mean pitch, and pitch variation – in a separate linear mixed effects model (features were centered). Fixed effects included Cognitive Status (cognitively normal (CN), MCI, dementia; treatment coded, ref = CN), Command Category (IOT, reminder, question; sum coded), and their interaction. Fixed effects also included Proportion of Experiment (centered) and speaker demographics: Age (centered), Gender (female, male; sum coded), Race (white, Black/African American, Asian; sum coded), Education (less than high school < high school degree < some college < college degree < professional degree; ordered factor), and Prior Voice Technology Experience (yes, no; sum coded). Random effects included by-Speaker and by-Trial random intercepts, and by-Speaker random slopes for Command Category and Proportion of the Experiment. In cases of singularity or convergence errors, the random effects structure was simplified [6].

All of the models had low collinearity (VIF < 5) and thus the predictors could be retained (outputs provided in **Appendices D, E, F, G, H, I, J, and K**). Several models showed effects of Cognitive Status: as seen in **Figure 2.**, relative to cognitively normal adults, adults with dementia produced a slower speech rate (*Coef* = -0.25, *t* = -2.87, *p* < 0.01), an overall slower articulation rate (exclusive of pauses) (*Coef* = -0.74, *t* = -7.46, *p* < 0.001), decreased pause ratio (*Coef* = -0.08, *t* = -4.01, *p* <0.001), decreased intensity (*Coef* = -1.54, *t* = -5.96, *p* < 0.001), decreased mean pitch (*Coef* = -1.93, *t* = -2.78, *p* <0.01), but with increased shimmer (*Coef* = 0.47, *t* = 2.82, *p* < 0.01). The pitch variation and jitter models showed no differences by Cognitive Status.

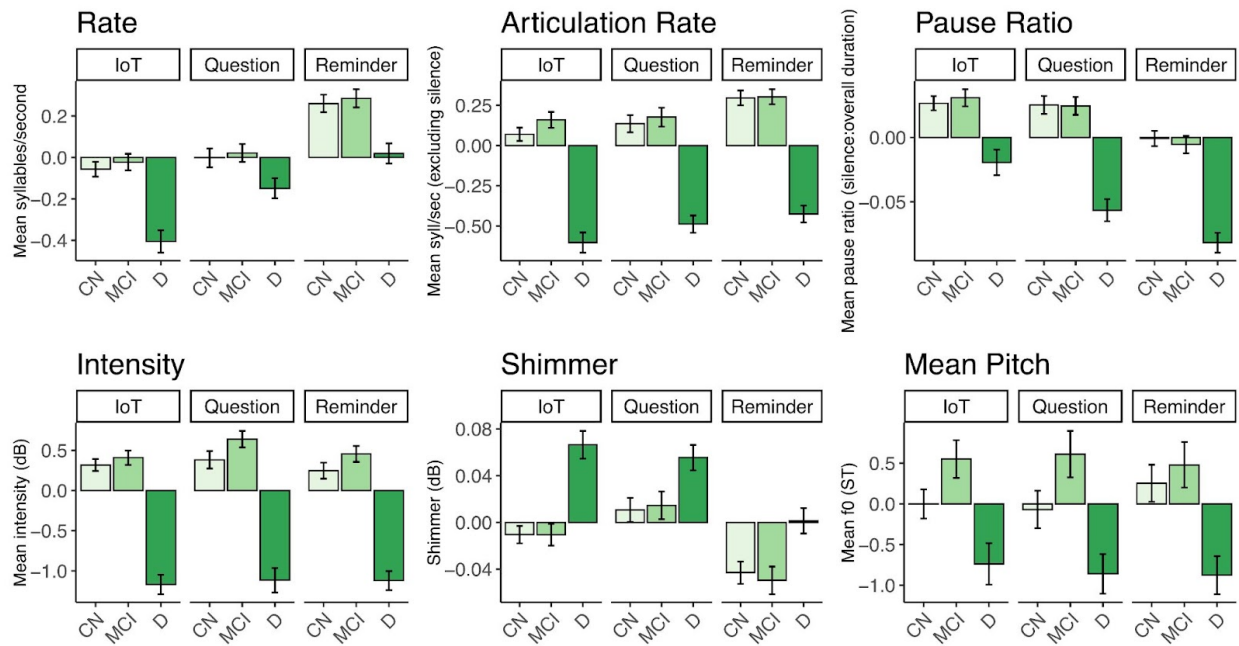

**Figure 2.** Acoustic measurements for speech rate, articulation rate, pause ratio, intensity, shimmer, and mean pitch across the groups: cognitively normal (CN), adults with mild cognitive impairment (MCI), and adults with dementia (D). Voice commands were produced across three categories: Internet of Things (IoT), Questions, and Reminders. Error bars indicate standard error of the mean.

**Acoustic features and ASR Accuracy**
To determine whether these speech differences with cognitive status reliably predict ASR accuracy, we modeled WER with a mixed effects linear regression, with the predictors (all centered and scaled) of speech rate, articulation rate, pause ratio, intensity, jitter, shimmer, mean pitch, pitch variation and all two-way interactions with ASR Model. The models also included the demographic covariates (gender, age, race, education, prior voice assistant experience). Random effects included by-Speaker and by-Trial random intercepts. The model showed high collinearity between speech rate and articulation rate. We created two models, one without speech rate and the other without articulation rate, but the rest of the main model structure. We assessed model comparisons with the AICc using the *MuMiN* R package [7]; the model excluding articulation rate (including speech rate) had better model fit (ΔAICc =-90.64) and had low collinearity (VIF < 5).

The retained model (output provided in **Appendix L**) showed that a higher WER was associated with greater shimmer (*Coef* = 0.01, *t* = 2.13, *p* < 0.05), lower intensity (*Coef* = -0.01, *t* = -3.01, *p* < 0.01), and a lower pause rato (*Coef* = -0.02, *t* = -4.96, *p* <0.001). There were four interactions between features and ASR Model: for the 'medium' model, WER increased with greater intensity (*Coef* = 0.01, *t* = 2.29, *p* < 0.05), speech rate (*Coef* = 0.01, *t* = 2.73, *p* < 0.01), and pause ratio (*Coef* = 0.01, *t* = 2.95, *p* < 0.01), but decreased with greater jitter (*Coef* = -0.01, *t* = -2.39, *p* <0.05). Put another way, as ASR Model was sum coded, the 'small' model had a higher WER for productions with greater jitter, a slower rate, lower intensity, and lower pause ratio.

**Discussion**

This paper conducted a quantitative analysis of ASR performance for older adults with and without cognitive impairment, examining voice commands in the VAS corpus [46]. These voice commands were pre-scripted sentences that speakers read aloud (e.g., "Alexa, tell me my reminders."). This design allows us to hold the content constant, as well as control for specific lexical and syntactic choices which otherwise vary with cognitive impairment [28, 34] (for a review, see [12].

     Our first hypothesis was that ASR accuracy would be lower for individuals with MCI and dementia, relative to cognitively normal adults. Indeed, we found evidence for this: adults with dementia had an error rate averaging 0.32 that was more than double that of cognitively normal adults (averaging ~ 0.12). This is a difference that is consistent even when taking into account task-related factors as well as speakers' demographic characteristics. This reduced ASR accuracy is consistent with prior work reporting descriptive statistics of higher WER for individuals with cognitive impairment [77]. However, we did not observe differences between adults with no cognitive impairment and those with MCI (WERs averaging ~0.12 and ~0.14, respectively). This contrasts with related findings of reduced accuracy (e.g., [46]), but might indicate that other factors could be leading to the reduced ASR accuracy for individuals in other work. An alternative explanation is that the task analyzed in the present study– reading pre-scripted commands – might have obscured differences that could emerge between cognitively normal and adults with MCI in spontaneous speech (e.g., [55, 80]).

     Our analysis additionally examined ASR performance across categories of commands. We found that ASR accuracy is especially reduced for adults with dementia in IoT commands (e.g., "Alexa, turn on the bedroom lights."), but is roughly equivalent to adults with no impairment and MCI for reminders. These decreases for IoT tasks can pose difficulty for aging-in-place systems aimed at supporting older adults with dementia; our findings suggest that they show even larger difficulties being understood by ASR when they are trying to control other smart objects. Recall that all of the utterances were pre-scripted and those analyzed in this ASR accuracy analysis were those that matched the commands. We anticipate that this gap

will be much wider for spontaneous speech and when individuals with dementia might be trying to control a smart object.

To determine whether differences in ASR accuracy are consistent across models, we compared two Whisper models: 'small', 'medium'. Both models are relatively lightweight and can be run locally on a laptop, which allows for HIPAA compliant processing. We found that while the larger model (here, Whisper 'medium') reduces word error rate (WER) for all speakers, this improvement was stronger for individuals with dementia.

Our second hypothesis was that acoustic features related to cognitive impairment – including timing, voice quality, and prosodic features – could predict ASR accuracy. To our knowledge, this is the first analysis of ASR performance based on acoustic properties of speech across individuals varying in cognitive status. Our results showed that individuals with dementia produced slower speech (both overall speech rate and articulation rate), lower pitched and quieter utterances, with greater shimmer, consistent with related work [3, 28, 54, 58]. While increased pauses have been linked to cognitive impairment in both reading [54] and producing spontaneous speech [28], we found that individuals with dementia actually produced a *lower* pause ratio. One possible explanation for this is the nature of the task: speakers read a voice command that included the wake word ("Alexa"), after which they needed to pause. Although beyond the scope of the current study, future research could examine whether the reduced pause ratio stems from a weaker pause between wake word and command.

When connecting the acoustic features to ASR accuracy, only a subset of the properties were predictive: greater shimmer, lower intensity, and lower pause ratio. These were the three acoustic features associated with individuals with dementia and led to consistently lower ASR accuracy. Prior work has shown that speakers with voice quality issues, including increased shimmer and jitter, show reduced ASR accuracy [31]. Reduced intensity in conversational speech has also been associated with lower ASR accuracy [82]. Furthermore, a slower speech rate (e.g., by manipulating recordings at 25% of their original rate) reduces ASR accuracy, extending pauses [64], aligning with our finding that a lower pause ratio was associated with reduced WER.

**HCI Implications for AgeTech & Voice Interfaces**
Our findings show that ASR errors affect accessibility for older adults with dementia -- the very population many AgeTech systems aim to support. These failures have direct implications for HCI: recognition errors can erode user trust [9], reduce satisfaction [59], and lead to technology abandonment [16, 50]. To address these challenges, we outline the following design suggestions:

**1. Speaker-personalized ASR models for ability-based adaptation.** We propose enabling ASR adaptation through two approaches: (1) fine-tuning ASR models on speech from older adults with cognitive impairment and (2) personalizing models to individual users using brief calibration tasks (e.g., reading short phrase sets) during initial setup and periodically thereafter for personalization. Our acoustic analyses show that dementia-related speech patterns diverge from those in large pretrained corpora, contributing to degraded ASR performance and motivating fine-tuning that adapts models to speech features observed with cognitive impairment (e.g., slowed rate, decreased intensity). We first propose fine-tuning ASR models on speech from the VAS corpus [46] to better capture dementia-related speech patterns, after which speaker-specific personalization can then be applied incrementally as the user interacts with the system.

As cognitive decline involves gradual within-speaker change, we argue for incremental, speaker-specific adaptation aligned with Ability-Based Design (Wobbrock et al., 2011). Prior work shows that speaker-specific adaptation reduces WER for speakers with atypical speech, mitigating disparities in speaker-independent models [27, 82, 83]. Personalization can leverage read calibration phrases, corrected commands, and high-confidence system outputs. Lightweight on-device adaptation (e.g., updating speaker embeddings) can preserve privacy, whereas server-side adaptation may enable more substantial model updates at the cost of transmitting audio or derived features [30, 33]. Personalization may involve fine-tuning a subset of model parameters rather than the full model, balancing performance gains with computational and energy constraints [53, 78] (Magister et al., 2025; Tahir et al., 2026). These design choices have direct implications for privacy, transparency, and user trust, particularly for older adults and caregivers.

**2. Human-in-the-loop personalization with users and caregivers.** We propose extending speaker personalization through a human-in-the-loop system that enables users and authorized caregivers to asynchronously review and correct ASR transcriptions paired with audio. While error detection and repair have been studied for speech interfaces [32, 36], many users struggle to identify errors in real time [32]. For users with cognitive impairment, this challenge is exacerbated: errors may go unnoticed and multi-turn repair sequences increase cognitive load. This motivates shifting repair work away from real-time interaction and toward asynchronous, low-pressure review.

Candidate utterances flagged by low-confidence estimates can be surfaced for review and validated corrections can serve as input for speaker-specific adaptation. This approach aligns with prior work where hearing collaborators supported users who were deaf and hard-of-hearing by correcting of ASR transcriptions, which reduced WER and improved user perceptions of understandability [40, 41, 96]. Providing users or caregivers with the original recording can support error detection, as recognition errors are difficult to identify in an unfamiliar voice [32]. Framing transcription review as "helping the system understand you or your loved one" positions the system as cooperative and caregivers may be motivated to review transcriptions when their corrections improve support for activities of daily living (e.g., setting reminders, controlling the home). Critically, giving users control over when and whether caregivers can access recordings supports their autonomy and privacy. This model aligns with a distributed cognition perspective, in which system intelligence emerges through collaboration between people and technology, rather than from automation alone.

**3. Adapting system timing and turn-taking to users' speech patterns**
Beyond model-level adaptation, our findings highlight the importance of interaction-level personalization (see also [17, 18]). Speakers with dementia exhibited slower speech rates, suggesting the need for delayed end-of-utterance detection and adaptive turn-taking strategies that reduce pressure to respond quickly. Quieter speech may similarly motivate dynamic microphone gain and explicit feedback when the system is listening. Together, these adaptations move beyond improving transcription accuracy alone, reshaping interaction timing and feedback to better align with users' speech production patterns.

**4. Moving beyond the wake-word.** Recognition accuracy is not the only interaction barrier faced by adults with cognitive impairment. In our study, two participants with dementia produced very few correctly recorded intents due to difficulty producing the correct wake word, leading to their exclusion from our ASR accuracy analysis. Consistent with prior recommendations [17], we suggest supporting alternative engagement mechanisms alongside wake words, such as a physical or on-screen button to initiate interaction. For example, [65] describe a button-based interaction for older adults with cognitive impairment that allows the user to control when recording starts and ends. These designs underscore the importance of providing redundant, accessible pathways for engagement, allowing users to select interaction modes that best match their current abilities.

Together, these implications emphasize voice interfaces that adapt across abilities, rather than assuming stable speech, memory, or interactional norms. Although grounded in AgeTech contexts, these strategies may generalize to other populations that experience persistent ASR barriers due to a mismatch between system assumptions and users' speech patterns (e.g., Down Syndrome [15], dysarthria [70]).

**Limitations and future directions**
This study has several limitations that can serve as avenues for future research. First, the recordings come from a corpus of read voice commands. While this affords experimental control — and well controlled comparison between the speakers — there are other linguistic features that differ with cognitive impairment (e.g. lexical, syntactic) that could also lead to ASR and NLU problems in spontaneous speech. Indeed, Whisper is more accurate at transcribing read speech than spontaneous speech [24, 25, 83]. Therefore, we might predict that some of the differences we observed between adults with no cognitive impairment and those with dementia might be magnified in spontaneous utterances. Furthermore, we predict that WER could be higher for adults with MCI in spontaneous speech tasks. As our analyses rely on scripted utterances, the present findings should be interpreted as a conservative estimate of recognition challenges in real-world voice interactions. This limits direct generalization of absolute WER values, but strengthens the motivation for the adaptive interaction strategies proposed here, which are likely to become even more critical in naturalistic use.

Second, we examined technology-directed utterances. While this type of speech aligns closely with the use cases we are most interested in, in particular smart home and voice assistant systems to support aging-in-place (e.g., [20, 39, 72]), it is possible that some of the technology-directed adaptations speakers make could have obscured differences between CN and MCI. This is potentially the case for acoustic features as some of these features are the same ones that people adjust when talking to technology (e.g., intensity, rate). Therefore they might be exaggerated or obscure differences.

Third, the data in the VAS corpus consisted of individuals with clinical diagnosis of dementia while those in the MCI group were made based on MoCA scores [42, 46]. This might have obscured differences between MCI and cognitively normal groups. Future work examining individuals with clinical diagnoses of MCI -- as well as MCI subtypes, such as amnestic and non-amnestic mild cognitive impairment (aMCI and naMCI, respectively) are needed for a more complete picture of the effect of MCI on speech and ASR outcomes. Relatedly, the VAS corpus did not specify a subtype of dementia (e.g., Alzheimer's Disease Related Dementia, Parkinson's, Vascular, Frontotemporal, Primary Progressive Aphasia), which can also lead to different patterns of linguistic features [2, 89].

Finally, this study consisted of just one language (English) and cultural context (United States). The acoustic properties we linked to cognitive impairment might not hold when comparing with other varieties of English and other languages, which can serve as an avenue for future research. Additionally, English is considered a 'high-resource' language; these are languages that have a large amount of training data to create robust language models (e.g., Mandarin Chinese, Spanish, French) [22, 29, 69]. However, the rest of the world's languages do not have the same degree of support and are considered 'low resource' languages (e.g., Cantonese). While there is much work to bridge these gaps [69, 97], speaking an 'under-resourced' language could pose an additional barrier for older adults seeking technology support. Future studies examining other high- and low-resource languages are needed to fully elucidate the effect of cognitive impairment on ASR performance.

**Conclusion**
Overall, this study shows that ASR accuracy is significantly worse in voice commands produced by older adults with dementia than cognitively normal adults. Furthermore, we found that ASR accuracy is especially reduced for adults with dementia in IoT commands (e.g., "Alexa, turn on the bedroom lights."). We additionally conducted an acoustic analysis of the voice commands, where we found that intensity, shimmer, and pause ratio were related to WER. Taken together, these findings point to concrete opportunities for designing more inclusive voice-based AgeTech. In particular, personalizing ASR models with speech samples from individuals with cognitive impairment may improve recognition and reduce ability-related disparities. More broadly, our results underscore the importance of designing voice interfaces that adapt to users' cognitive and speech-production abilities, supporting equitable access.

**Acknowledgments**
This work was supported by a UC Davis Academic Federation Research Grant to A.W., C.C., G.Z., and M.C.

**Appendix A. Voice assistant commands**

| Category | Trial | Command |
| --- | --- | --- |
| Question | 1 | Alexa, what is the weather outside? |
| Question | 2 | Alexa, what is today's date? |
| Question | 3 | Alexa, what time is it? |
| Question | 4 | Alexa, when is Thanksgiving? |
| Question | 5 | Alexa, how do you bake chocolate chip cookies? |
| Question | 6 | Alexa, what is 2 times 4? |
| Question | 7 | Alexa, how many tablespoons in a cup? |
| Question | 8 | Alexa, how do you spell 'symptom'? |
| IoT | 9 | Alexa, play classical music. |
| IoT | 10 | Alexa, volume 8. |
| IoT | 11 | Alexa, play Jazz. |
| IoT | 12 | Alexa, volume 6. |
| Reminder | 13 | Alexa, rebmind me to start the laundry tomorrow at 2pm. |
| Reminder | 14 | Alexa, remind me to feed the dog at 7pm everyday. |
| Reminder | 15 | Alexa, tell me my reminders. |
| Reminder | 16 | Alexa, remember my daughter's birthday is June first. |
| Reminder | 17 | Alexa, set a timer in 5 seconds. |
| Reminder | 18 | Alexa, set my alarm for 7am tomorrow. |
| Reminder | 19 | Alexa, add oranges and grapes to my shopping list. |
| Reminder | 20 | Alexa, what is in my shopping list? |
| IoT | 21 | Alexa, call (603)660-2203. |
| IoT | 22 | Alexa, find my phone. |
| IoT | 23 | Alexa, turn the bedroom light on. |
| IoT | 24 | Alexa, turn the bedroom light red. |
| IoT | 25 | Alexa, change brightness to 10. |
| IoT | 26 | Alexa, turn off the bedroom light. |
| IoT | 27 | Alexa, open the kitchen camera. |
| IoT | 28 | Alex, hide the kitchen camera. |
| IoT | 29 | Alexa, play White Collar on Fire TV |
| IoT | 30 | Alexa, pause |

**Appendix B.**
We ran the utterances through the models using the default Whisper values
- compression_ratio_threshold = 2.4
- logprob_threshold = -1.0
- no_speech_threshold = 0.6
- temperature = (0.0, 0.2, 0.4, 0.6, 0.8, 1.0)

with an additional beam size=5 to increase transcription accuracy with fewer hallucinations.

## Appendix C. Word error rate (WER) model output

|  | Coef | Std. Error | df | t value | p value |
|---|---|---|---|---|---|
| (Intercept) | -0.01 | 0.03 | 90.97 | -0.26 | 0.79 |
| CognitiveStatus(MCI) | 3.7e-03 | 0.02 | 68.66 | 0.16 | 0.87 |
| **CognitiveStatus(D)** | **0.14** | **0.03** | **71.81** | **5.17** | **<0.001** |
| CommandCategory(IoT) | 0.04 | 0.02 | 39.16 | 1.54 | 0.13 |
| CommandCategory(Q) | -0.03 | 0.03 | 43.24 | -1.12 | 0.27 |
| **ASR_Model(med)** | **-0.03** | **0.01** | **4476.08** | **-6.04** | **<0.001** |
| Proportion Experiment | -0.02 | 0.04 | 401.11 | -0.45 | 0.65 |
| Age | 1.4e-03 | 1.7e-03 | 71.15 | 0.79 | 0.43 |
| Gender(f) | -0.02 | 0.01 | 69.39 | -1.77 | 0.08 |
| Race(AfAm) | -0.04 | 0.03 | 69.07 | -1.21 | 0.23 |
| Race(white) | -0.04 | 0.02 | 68.39 | -1.61 | 0.11 |
| Education.L | 0.02 | 0.04 | 69.84 | 0.5 | 0.62 |
| Education.Q | -0.04 | 0.03 | 69.67 | -1.07 | 0.29 |
| Education.C | 0.02 | 0.02 | 69.52 | 1.03 | 0.31 |
| Prior Technology Experience (yes) | -0.01 | 0.01 | 69.23 | -0.73 | 0.47 |
| CognitiveStatus(MCI):CommandCategory(IoT) | -0.01 | 0.01 | 749.7 | -0.81 | 0.42 |
| **CognitiveStatus(D):CommandCategory(IoT)** | **0.11** | **0.01** | **1208.17** | **7.7** | **<0.001** |
| CognitiveStatus(MCI):CommandCategory(Q) | 0.02 | 0.01 | 514.3 | 1.56 | 0.12 |
| CognitiveStatus(D):CommandCategory(Q) | -0.02 | 0.01 | 472.29 | -1.15 | 0.25 |
| CognitiveStatus(MCI):ASR_Model(med) | -0.01 | 0.01 | 4476.11 | -1.21 | 0.23 |
| **CognitiveStatus(D):ASR_Model(med)** | **-0.08** | **0.01** | **4476.48** | **-8.84** | **<0.001** |
| **CommandCategory(IoT):ASR_Model(med)** | **-0.01** | **4.7e-03** | **4476.37** | **-2.78** | **<0.01** |
| **CommandCategory(Q):ASR_Model(med)** | **-0.01** | **0.01** | **4476.21** | **-2.01** | **0.04** |
| *Random effects* | *Variance* | | | | |
| Speaker (intercept) | 0.01 | | | | |
|    Proportion Experiment | 0.01 | | | | |
| Trial (intercept) | 0.01 | | | | |
| *Num. observations* = 4692, *num. Speakers* = 83, *num. Trials* = 28 | | | | | |

Releveled model showed an interaction between Command Category (Reminder) and Cognitive Status: a reduction in WER for individuals with dementia for reminders (\textit{Coef} = -0.08, \textit{t} = -6.76, \textit{p} \textless{} 0.001). Additionally, the releveled model showed an interaction between Command Category and ASR Model: less of a decrease of WER for the medium model for reminders (\textit{Coef} = 0.02, \textit{t} = 4.61, \textit{p} \textless{} 0.001).

**Appendix D. Speech Rate model output**

|  | Estimate | Std. Error | df | t value | p value |
|---|---|---|---|---|---|
| (Intercept) | -0.03 | 0.12 | 93.84 | -0.23 | 0.82 |
| CognitiveStatus(MCI) | 4.3e-03 | 0.07 | 71.3 | 0.06 | 0.95 |
| **CognitiveStatus(D)** | **-0.25** | **0.09** | **75.04** | **-2.87** | **< 0.01** |
| CommandCategory(IoT) | -0.13 | 0.08 | 39.73 | -1.54 | 0.13 |
| CommandCategory(Q) | -0.07 | 0.1 | 43.78 | -0.67 | 0.5 |
| Proportion Experiment | -1.8e-03 | 0.15 | 336.23 | -0.01 | 0.99 |
| Age | 2.3e-03 | 0.01 | 74.22 | 0.4 | 0.69 |
| **Gender(f)** | **0.08** | **0.03** | **72.13** | **2.52** | **0.01** |
| Race(AfAm) | 0.04 | 0.1 | 71.8 | 0.37 | 0.71 |
| Race(white) | 0.05 | 0.08 | 70.95 | 0.62 | 0.54 |
| Education.L | 0.18 | 0.13 | 72.49 | 1.38 | 0.17 |
| Education.Q | -0.1 | 0.11 | 72.3 | -0.93 | 0.36 |
| Education.C | 0.13 | 0.08 | 72.21 | 1.68 | 0.1 |
| Prior Technology Experience (yes) | 0.01 | 0.03 | 71.95 | 0.41 | 0.69 |
| CognitiveStatus(MCI):CommandCategory(IoT) | 0.01 | 0.04 | 583.14 | 0.15 | 0.88 |
| CognitiveStatus(D):CommandCategory(IoT) | -0.04 | 0.05 | 963.33 | -0.92 | 0.36 |
| CognitiveStatus(MCI):CommandCategory(Q) | -0.01 | 0.05 | 392.01 | -0.21 | 0.83 |
| CognitiveStatus(D):CommandCategory(Q) | 0.08 | 0.05 | 395.25 | 1.51 | 0.13 |
| *Random effects* | *Variance* | | | | |
| Speaker (intercept) | 0.06 | | | | |
| Proportion Experiment | 0.05 | | | | |
| Trial (intercept) | 0.08 | | | | |

*Num. observations* = 2353, *num. Speakers* = 83, *num. Trials* = 29

Releveled model. Effect of Command Category: Reminders were produced at a faster rate (*Coef* = 0.20, *t* = -2.57, *p* < 0.05)

## Appendix E. Articulation Rate model output

|  | Estimate | Std. Error | df | t value | p value |
|---|---|---|---|---|---|
| (Intercept) | 0.07 | 0.12 | 83.83 | 0.56 | 0.58 |
| CognitiveStatus(MCI) | 2.6e-03 | 0.08 | 69.99 | 0.03 | 0.98 |
| **CognitiveStatus(D)** | **-0.74** | **0.1** | **74.29** | **-7.46** | **<0.001** |
| CommandCategory(IoT) | -0.07 | 0.07 | 50.25 | -1.02 | 0.31 |
| CommandCategory(Q) | -0.06 | 0.08 | 53.92 | -0.75 | 0.45 |
| Proportion Experiment | -0.11 | 0.15 | 109.44 | -0.75 | 0.46 |
| Age | 2.9e-03 | 0.01 | 73.59 | 0.45 | 0.65 |
| **Gender(f)** | **0.11** | **0.04** | **71.06** | **3.03** | **<0.01** |
| Race(AfAm) | 0.08 | 0.11 | 70.63 | 0.73 | 0.47 |
| Race(white) | 0.01 | 0.09 | 69.61 | 0.14 | 0.89 |
| **Education.L** | **0.32** | **0.15** | **71.73** | **2.17** | **0.03** |
| Education.Q | -0.21 | 0.12 | 71.46 | -1.69 | 0.1 |
| Education.C | 0.14 | 0.09 | 71.23 | 1.64 | 0.1 |
| Prior Technology Experience (yes) | -0.05 | 0.04 | 70.88 | -1.51 | 0.14 |
| CognitiveStatus(MCI):CommandCategory(IoT) | 0.04 | 0.05 | 503.11 | 0.74 | 0.46 |
| CognitiveStatus(D):CommandCategory(IoT) | 0.01 | 0.06 | 902.79 | 0.09 | 0.93 |
| CognitiveStatus(MCI):CommandCategory(Q) | -4.6e-03 | 0.06 | 323.58 | -0.08 | 0.93 |
| CognitiveStatus(D):CommandCategory(Q) | 0.05 | 0.06 | 345.43 | 0.77 | 0.44 |
| *Random effects* | *Variance* | | | | |
| Speaker (intercept) | 0.07 | | | | |
|    Proportion Experiment | 0.02 | | | | |
| Trial (intercept) | 0.04 | | | | |
| *Num. observations = 2353, num. Speakers = 83, num. Trials = 29* | | | | | |

Releveled model. No effect of Command Category (Reminders) on articulation rate (*Coef* = 0.14, *t* = 1.96, *p* = 0.06).

**Appendix F. Pause Ratio model output**

|  | Estimate | Std. Error | df | t value | p value |
|---|---|---|---|---|---|
| (Intercept) | 0.02 | 0.02 | 81.89 | 0.89 | 0.38 |
| CognitiveStatus(MCI) | -2.0e-03 | 0.02 | 70.61 | -0.12 | 0.91 |
| **CognitiveStatus(D)** | **-0.08** | **0.02** | **72.34** | **-4.01** | **<0.001** |
| CommandCategory(IoT) | 0.01 | 0.01 | 40.1 | 1.27 | 0.21 |
| CommandCategory(Q) | 2.5e-03 | 0.01 | 44.39 | 0.18 | 0.86 |
| Proportion Experiment | -0.01 | 0.02 | 225.87 | -0.66 | 0.51 |
| Age | 2.0e-04 | 1.3e-03 | 71.97 | 0.16 | 0.88 |
| Gender(f) | -7.0e-04 | 0.01 | 71.05 | -0.1 | 0.92 |
| Race(AfAm) | 0.01 | 0.02 | 70.87 | 0.42 | 0.68 |
| Race(white) | -0.01 | 0.02 | 70.49 | -0.54 | 0.59 |
| Education.L | 0.01 | 0.03 | 71.26 | 0.5 | 0.62 |
| Education.Q | -0.01 | 0.03 | 71.16 | -0.46 | 0.65 |
| Education.C | -0.01 | 0.02 | 71.09 | -0.31 | 0.76 |
| Prior Technology Experience (yes) | -0.01 | 0.01 | 70.99 | -1.58 | 0.12 |
| CognitiveStatus(MCI):CommandCategory(IoT) | 2.8e-03 | 0.01 | 721.79 | 0.42 | 0.67 |
| CognitiveStatus(D):CommandCategory(IoT) | 0.01 | 0.01 | 1058.15 | 1.38 | 0.17 |
| CognitiveStatus(MCI):CommandCategory(Q) | 2.2e-03 | 0.01 | 512.84 | 0.28 | 0.78 |
| CognitiveStatus(D):CommandCategory(Q) | -0.01 | 0.01 | 479.74 | -0.72 | 0.47 |
| *Random effects* | *Variance* | | | | |
| Speaker (intercept) | 3.4e-03 | | | | |
|    Proportion Experiment | 2.5e-03 | | | | |
| Trial (intercept) | 0.01 | | | | |
| *Num. observations* = 2353, *num. Speakers* = 83, *num. Trials* = 29 | | | | | |

Releveled model showed no effects of Command Category.

**Appendix G. Intensity model output**

|  | Estimate | Std. Error | df | t value | p value |
|---|---|---|---|---|---|
| (Intercept) | 0.47 | 0.31 | 78.98 | 1.5 | 0.14 |
| CognitiveStatus(MCI) | 0.04 | 0.22 | 70.91 | 0.19 | 0.85 |
| **CognitiveStatus(D)** | **-1.54** | **0.26** | **72.92** | **-5.96** | **<0.001** |
| CommandCategory(IoT) | 0.16 | 0.14 | 50.18 | 1.12 | 0.27 |
| CommandCategory(Q) | -0.13 | 0.17 | 54.49 | -0.79 | 0.43 |
| Proportion Experiment | -0.71 | 0.31 | 120.73 | -2.3 | 0.02 |
| Age | 0.02 | 0.02 | 72.21 | 0.93 | 0.36 |
| Gender(f) | 0.1 | 0.09 | 70.75 | 1.03 | 0.31 |
| Race(AfAm) | -0.18 | 0.3 | 70.57 | -0.61 | 0.54 |
| Race(white) | -0.16 | 0.24 | 69.95 | -0.67 | 0.5 |
| Education.L | 0.15 | 0.39 | 70.66 | 0.39 | 0.7 |
| Education.Q | -0.11 | 0.32 | 70.58 | -0.35 | 0.73 |
| Education.C | 0.46 | 0.22 | 70.7 | 2.09 | 0.04 |
| Prior Technology Experience (yes) | -0.01 | 0.09 | 70.56 | -0.12 | 0.9 |
| CognitiveStatus(MCI):CommandCategory(IoT) | -0.03 | 0.11 | 755.21 | -0.27 | 0.78 |
| CognitiveStatus(D):CommandCategory(IoT) | 0.08 | 0.12 | 1075.02 | 0.71 | 0.48 |
| CognitiveStatus(MCI):CommandCategory(Q) | -0.02 | 0.12 | 543.78 | -0.16 | 0.88 |
| CognitiveStatus(D):CommandCategory(Q) | -0.11 | 0.13 | 498.58 | -0.87 | 0.39 |
| *Random effects* | *Variance* | | | | |
| Speaker (intercept) | 0.52 | | | | |
|    Proportion Experiment | 0.72 | | | | |
| Trial (intercept) | 0.16 | | | | |

*Num. observations* = 2353, *num. Speakers* = 83, *num. Trials* = 29

The releveled model showed no effects of Command Category.

**Appendix H. Jitter model output**

|  | Estimate | Std. Error | df | t value | p value |
|---|---|---|---|---|---|
| (Intercept) | -0.29 | 0.22 | 75.51 | -1.34 | 0.19 |
| CognitiveStatus(MCI) | 0.15 | 0.16 | 71.52 | 0.93 | 0.36 |
| CognitiveStatus(D) | 0.25 | 0.19 | 72.41 | 1.33 | 0.19 |
| CommandCategory(IoT) | -2.4e-03 | 0.06 | 48.22 | -0.04 | 0.97 |
| **CommandCategory(Q)** | **0.16** | **0.07** | **53.09** | **2.15** | **0.04** |
| Proportion Experiment | -0.03 | 0.13 | 162.17 | -0.23 | 0.82 |
| Age | -0.01 | 0.01 | 71.96 | -1.22 | 0.23 |
| **Gender(f)** | **-0.59** | **0.07** | **71.48** | **-8.88** | **<0.001** |
| Race(AfAm) | -0.1 | 0.22 | 71.37 | -0.46 | 0.64 |
| Race(white) | 0.12 | 0.17 | 71.17 | 0.71 | 0.48 |
| Education.L | 0.24 | 0.28 | 71.65 | 0.87 | 0.39 |
| Education.Q | -0.17 | 0.23 | 71.58 | -0.72 | 0.48 |
| Education.C | 0.02 | 0.16 | 71.51 | 0.13 | 0.89 |
| Prior Technology Experience (yes) | 0.03 | 0.07 | 71.46 | 0.46 | 0.65 |
| **CognitiveStatus(MCI):CommandCategory(IoT)** | **0.09** | **0.04** | **799.75** | **2.02** | **0.04** |
| CognitiveStatus(D):CommandCategory(IoT) | 0.09 | 0.05 | 1094.49 | 1.84 | 0.07 |
| CognitiveStatus(MCI):CommandCategory(Q) | -4.6e-03 | 0.05 | 585.43 | -0.09 | 0.93 |
| CognitiveStatus(D):CommandCategory(Q) | -2.2e-03 | 0.05 | 521.67 | -0.04 | 0.97 |
| *Random effects* | *Variance* | | | | |
| Speaker (intercept) | 0.30 | | | | |
|    Proportion Experiment | 0.14 | | | | |
| Trial (intercept) | 0.04 | | | | |
| *Num. observations = 2353, num. Speakers = 83, num. Trials = 29* | | | | | |

Releveled model. Effect of Command Category: Reminders were produced with less jitter (*Coef* = -0.16, *t* = 2.27, *p* < 0.05)

**Appendix I. Shimmer model output**

|  | Estimate | Std. Error | df | t value | p value |
|---|---|---|---|---|---|
| (Intercept) | -0.05 | 0.2 | 79.94 | -0.23 | 0.82 |
| CognitiveStatus(MCI) | 0.08 | 0.14 | 71.88 | 0.56 | 0.58 |
| **CognitiveStatus(D)** | **0.47** | **0.17** | **73.23** | **2.82** | **<0.01** |
| CommandCategory(IoT) | -0.09 | 0.08 | 45.18 | -1.11 | 0.27 |
| **CommandCategory(Q)** | **0.29** | **0.1** | **49.62** | **2.98** | **<0.01** |
| **Proportion Experiment** | **0.48** | **0.17** | **165.26** | **2.85** | **<0.01** |
| Age | -1.8e-03 | 0.01 | 72.71 | -0.16 | 0.87 |
| **Gender(f)** | **-0.31** | **0.06** | **71.76** | **-5.25** | **<0.001** |
| Race(AfAm) | -0.25 | 0.19 | 71.65 | -1.28 | 0.2 |
| Race(white) | -0.16 | 0.16 | 71.24 | -1.02 | 0.31 |
| Education.L | 0.2 | 0.25 | 71.66 | 0.81 | 0.42 |
| Education.Q | -0.11 | 0.21 | 71.62 | -0.52 | 0.6 |
| Education.C | -0.06 | 0.14 | 71.72 | -0.41 | 0.68 |
| Prior Technology Experience (yes) | 0.09 | 0.06 | 71.63 | 1.44 | 0.16 |
| CognitiveStatus(MCI):CommandCategory(IoT) | 0.02 | 0.05 | 651.42 | 0.32 | 0.75 |
| CognitiveStatus(D):CommandCategory(IoT) | 0.11 | 0.06 | 1015.56 | 1.82 | 0.07 |
| CognitiveStatus(MCI):CommandCategory(Q) | 0.05 | 0.06 | 449.48 | 0.75 | 0.45 |
| CognitiveStatus(D):CommandCategory(Q) | -0.02 | 0.06 | 438.23 | -0.32 | 0.75 |
| *Random effects* | *Variance* | | | | |
| Speaker (intercept) | 0.23 | | | | |
| Proportion Experiment | 0.10 | | | | |
| Trial (intercept) | 0.07 | | | | |
| *Num. observations = 2353, num. Speakers = 83, num. Trials = 29* | | | | | |

Releveled model. Effect of Command Category: Reminders were produced with less shimmer (*Coef* = -0.20 *t* = -2.42, *p* < 0.05)

## Appendix J. Mean pitch model output

| | Estimate | Std. Error | df | t value | p value |
|---|---|---|---|---|---|
| (Intercept) | 1.14 | 0.8 | 72.1 | 1.42 | 0.16 |
| CognitiveStatus(MCI) | -0.03 | 0.59 | 69.31 | -0.05 | 0.96 |
| **CognitiveStatus(D)** | **-1.93** | **0.69** | **71.74** | **-2.78** | **< 0.01** |
| CommandCategory(IoT) | -0.17 | 0.1 | 53.99 | -1.7 | 0.09 |
| CommandCategory(Q) | 0.07 | 0.12 | 64.65 | 0.53 | 0.59 |
| Proportion Experiment | 0.41 | 0.23 | 77.18 | 1.81 | 0.07 |
| Age | 0.05 | 0.04 | 70.91 | 1.1 | 0.28 |
| **Gender(f)** | **3.04** | **0.24** | **69.92** | **12.43** | **<0.001** |
| Race(AfAm) | -0.35 | 0.8 | 69.91 | -0.44 | 0.66 |
| Race(white) | -1.22 | 0.64 | 69.15 | -1.91 | 0.06 |
| Education.L | 1.19 | 1.03 | 70.23 | 1.16 | 0.25 |
| Education.Q | -0.18 | 0.86 | 70.12 | -0.21 | 0.84 |
| Education.C | 0.19 | 0.6 | 69.98 | 0.32 | 0.75 |
| Prior Technology Experience (yes) | -0.05 | 0.25 | 69.86 | -0.22 | 0.83 |
| CognitiveStatus(MCI):CommandCategory(IoT) | -0.01 | 0.09 | 66.17 | -0.16 | 0.87 |
| CognitiveStatus(D):CommandCategory(IoT) | 0.17 | 0.1 | 85.14 | 1.64 | 0.1 |
| CognitiveStatus(MCI):CommandCategory(Q) | 0.14 | 0.12 | 77.33 | 1.17 | 0.24 |
| CognitiveStatus(D):CommandCategory(Q) | -2.9e-03 | 0.13 | 84.15 | -0.02 | 0.98 |
| **Age*Gender(f)** | **-0.08** | **0.04** | **71.03** | **-2.08** | **0.04** |
| *Random effects* | *Variance* | | | | |
| Speaker (intercept) | 4.77 | | | | |
|    Proportion Experiment | 0.81 | | | | |
|    Command Category(IoT) | 0.01 | | | | |
|    Command Category(Q) | 0.06 | | | | |
| Trial (intercept) | 0.06 | | | | |

*Num. observations* = 2353, *num. Speakers* = 83, *num. Trials* = 29

Releveled model showed no effects of Command Category.

**Appendix K. Pitch variation model**

|  | Estimate | Std. Error | df | t value | Pr(>|t|) |
|---|---|---|---|---|---|
| (Intercept) | -0.01 | 0.11 | 72.33 | -0.09 | 0.93 |
| CognitiveStatus(MCI) | 0.03 | 0.08 | 70.45 | 0.44 | 0.66 |
| CognitiveStatus(D) | 0.08 | 0.09 | 72.93 | 0.85 | 0.40 |
| CommandCategory(IoT) | 0.03 | 0.03 | 32.28 | 0.92 | 0.36 |
| CommandCategory(Q) | -0.02 | 0.03 | 32.54 | -0.74 | 0.46 |
| Proportion Experiment | -0.05 | 0.07 | 44.5 | -0.73 | 0.47 |
| Age | 1.0e-03 | 0.01 | 72.97 | 0.17 | 0.87 |
| **Gender(f)** | 0.02 | 0.03 | 71.48 | 0.48 | 0.64 |
| Race(AfAm) | -0.08 | 0.11 | 71.25 | -0.78 | 0.44 |
| Race(white) | -0.07 | 0.09 | 70.64 | -0.81 | 0.42 |
| Education.L | 0.13 | 0.14 | 71.65 | 0.93 | 0.36 |
| Education.Q | -0.04 | 0.12 | 71.53 | -0.37 | 0.71 |
| Education.C | -2.7e-03 | 0.08 | 71.51 | -0.03 | 0.97 |
| Prior Technology Experience (yes) | 8.5e-04 | 0.03 | 71.35 | 0.02 | 0.98 |
| *Random effects* | *Variance* | | | | |
| Speaker (intercept) | 0.01 | | | | |
| Proportion Experiment | 8.2e-05 | | | | |
| Trial (intercept) | 3.0e-03 | | | | |
| *Num. observations* =4692, *num. Speakers* = 83, *num. Trials* = 29 | | | | | |

Releveled model showed no effects of Command Category.
Note that including the interaction between Cognitive Status and Command Category led to high collinearity.

## Appendix L. Acoustics predicting ASR accuracy model output

|  | Estimate | Std. Error | df | t value | p value |
|---|---|---|---|---|---|
| (Intercept) | 0.06 | 0.03 | 89.07 | 1.87 | 0.07 |
| Rate | 2.6e-03 | 4.7e-03 | 4635.26 | 0.55 | 0.58 |
| Jitter | 0.01 | 0.01 | 4370.54 | 1.84 | 0.07 |
| **Shimmer** | **0.01** | **0.01** | **4526.3** | **2.13** | **0.03** |
| **Intensity** | **-0.01** | **4.5e-03** | **4418.69** | **-3.01** | **<0.01** |
| **Pause Ratio** | **-0.02** | **4.9e-03** | **3943.69** | **-4.96** | **<0.001** |
| Pitch Variation | -2.0e-03 | 3.9e-03 | 4561.44 | -0.51 | 0.61 |
| Mean Pitch | 0.01 | 0.01 | 795.39 | 0.88 | 0.38 |
| **ASR Model(med)** | **-0.06** | **3.5e-03** | **4554.67** | **-16.82** | **<0.001** |
| Proportion Experiment | -0.02 | 0.04 | 156.53 | -0.59 | 0.55 |
| **Age** | **0.01** | **1.6e-03** | **71.19** | **3.52** | **<0.001** |
| Gender(f) | -0.01 | 0.01 | 117.85 | -0.63 | 0.53 |
| Race(AfAm) | -6.1e-04 | 0.03 | 68.44 | -0.02 | 0.98 |
| **Race(white)** | **-0.06** | **0.03** | **67.3** | **-2.37** | **0.02** |
| Education.L | -0.01 | 0.04 | 69.21 | -0.22 | 0.83 |
| Education.Q | -0.02 | 0.03 | 68.09 | -0.66 | 0.51 |
| Education.C | 0.03 | 0.02 | 68.27 | 1.08 | 0.28 |
| Prior Technology Experience (yes) | -0.01 | 0.01 | 68.23 | -0.72 | 0.48 |
| **Rate*ASR Model (med)** | **0.01** | **4.1e-03** | **4554.66** | **2.73** | **<0.01** |
| **Jitter*ASR Model (med)** | **-0.01** | **0.01** | **4554.68** | **-2.39** | **0.02** |
| Shimmer*ASR Model (med) | -0.01 | 4.6e-03 | 4554.66 | -1.87 | 0.06 |
| **Intensity*ASR Model (med)** | **0.01** | **3.9e-03** | **4554.66** | **2.29** | **0.02** |
| **PauseRatio*ASR Model (med)** | **0.01** | **4.0e-03** | **4554.66** | **2.95** | **<0.01** |
| Pitch Variation*ASR Model (med) | 2.7e-03 | 3.5e-03 | 4554.66 | 0.77 | 0.44 |
| Mean Pitch*ASR Model (med) | -0.01 | 0.01 | 4554.65 | -1.49 | 0.14 |
| *Random effects* | Variance | | | | |
| Speaker (intercept) | 0.01 | | | | |
| Trial (intercept) | 0.01 | | | | |

*Num. observations* =4692, *num. Speakers* = 83, *num. Trials* = 29